\documentclass[runningheads,a4paper]{llncs}

 \usepackage{float}
\usepackage{graphicx}
\usepackage{mathrsfs}
\usepackage{latexsym}
\usepackage{amssymb}
\usepackage{amscd}
\usepackage{amsmath}
\usepackage{bbm}
\usepackage{url}
\usepackage{float}

\newcommand{\keywords}[1]{\par\addvspace\baselineskip
\noindent\keywordname\enspace\ignorespaces#1}

\begin{document}

\mainmatter 

\title{Is a `Wirikuta empowerment' of the Huichol measurable on the Internet?}
\titlerunning{ `Wirikuta empowerment' \& the Internet.}

\author{Lorena P\'erez Garc\'ia\thanks{These authors contributed equally to this work.}, Jan Broekaert$^\star$, Nicole Note}
\authorrunning{P\'erez Garc\'ia, Broekaert, Note}

\tocauthor{Lorena P\'erez Garc\'ia, Jan Broekaert, Nicole Note}

\institute{Center Leo Apostel for Interdisciplinary Studies, 
Brussels Free University,\\ Pleinlaan 2, 1050 Brussel, Belgium\\ \email{clperez\_garcia@hotmail.com,jbroekae@vub.ac.be,nnote@vub.ac.be}}


\toctitle{Is a `Wirikuta empowerment' measurable on the Internet?}

\maketitle

\begin{abstract}
Current social and activist movements find  the opportunity in social media to effectively impact on the agenda of governing bodies and create `global' perceptions -- it is often claimed. 
Content related to the social and activist movements is online, to be accessed, supported or disputed and distributed from virtually anywhere at any time, in the public sphere of the  Internet.
This activity  allows the enlargement of social movements and would increase the empowerment in the concerned communities.  
The aim of this explorative study is to assess whether the temporal evolution of the Normalised Web Distance (NWD) --as defined by  Cilibrasi \& Vit\'anyi  (2007)-- between identifying terms concerning this activism could be used to measure the progress or decline of social empowerment through the Internet. 
The NWD relies on the page count number of single and joint queries, which in our study have been registered using a freely available web browser (in this case Google Search) providing a time search window for temporal query results.
To explore this meta-data technique, we introduce the case of a perceived Wirikuta online movement, which originated in Mexico with the aim to protect the Huichols' sacred land and water resources from open mining projects for silver ore. 
We conducted a small scale Internet study relating the key terms \emph{Wirikuta\footnote{The alternative spelling \emph{Huiricuta} is less common, but appears in e.g. UNESCO documents.}, Huichol}, and \emph{Wixarika} and their co-occurrence with seven positive  qualifiers (e.g. \emph{sacred land}), five negative qualifiers (e.g. \emph{violence}) and one neutral qualifier (\emph{table})  over time, annually from 1994 till 2013. %
We find the accuracy of the temporal NWD-based method is limited due to --previously reported by e.g. Satoh and Yamana (2012)-- short-term variability and between-user variability of the search tool's page hit counts.  We confirm close semantic clustering  over time of traditional indigeneous identity terms of the Huichol, and observe a slight convergence of key terms to \emph{mines} and less pronounced to \emph{sacred land}  and a divergence with respect  to \emph{ancestors} indicating a complex image of a tendency of empowerment.  
\end{abstract}

\section{Introduction}
The Internet entices millions of users around the globe to  gain access to billions of web pages and social media entries, and actively create online content.
In this way the Internet constitutes to extent a self-organizing virtual and heterogenous public sphere \cite{rheingold1993,castells2008,freyermuth2010}.  Technologically and culturally,  instant communities of transformative practice emerge within this public space  \cite{castells2012}. 
Grass root social movements find in social media the opportunity to have an impact on both individuals and society without having to rely on either physical space or real-time, but crucially on ICT. 
The individual agents in this process co-create the dynamics of a social movement and potentially empower  --not necessarily their local--  social communities.
The complex entanglement of agonist and contragonist elements of online information, the wide spectrum and variety of interests, views and ideologies leads to an imbroglio of information which permanently changes over time. The majority of online autonomous resources is only meaningful in a low-quality approximate sense \cite{oliveretal1997}.  
Given the vast amount of information (data)  on the internet for a given subject, beyond the scope of individual perusal, could one still obtain meaningful information from meta-data about them? Would it be possible to measure evolutions of an empowerment --its reflection in the public sphere of the internet-- from a meta-data perspective? \\
We encountered this research problem when we were evaluating the empowerment in indigenous communities through ICT (e.g. \cite{kenkarasseril2013}), in Mexico \cite{thesisLPG2014}. We hypothesized the empowerment of `local' use of ICT could be reflected in global perception, forwarded by ICT. We focused on how a local, diaspora and sympathizing global `community' act to defend the Wirikuta, the Huichol's sacred land, from open-surface mining projects by international contractors. 
Our central research goal was to develop a method which could probe the Internet for empowerment achieved in communities in a synthetic or meta-level manner.  Thus not based on the perusal of individual informative online documents  --e.g. a governmental decision \cite{huauxa2008}-- but by an extensive evaluation of  `bulk' online content.\footnote{One could compare this approach to the measurement of temperature as a macroscopic property of bulk matter, emerging from the velocities of its individual particles.} Not only does this hypothesis suppose a strong relation between relevant deploying events and `same time' coverage by Internet content, it also heavily relies on a relation between semantic similarity of key terms and meaningful information. The latter relation is not only subject to interpretation but also to effects of syntax, metonymy, polysemy and language.\footnote{Our study mainly involves English terms given it is the \emph{lingua franca} of the Internet. The information transformation and flow of Wixarika  and Spanish data to English data is complication proper to global Internet dynamics. }\\
To establish our study (Sec.2) we  present our view on the Internet as a nexus of the public sphere\footnote{The troublesome percolation of  Habermass' `system' into the proper  lifeworld is considered reciprocally as well \cite{castells2008,habermas1992}.} allowing social movements to flourish and advance individual empowerment\footnote{The capability-approach of Sen and Nussbaum focuses on the acquired abilities of the individuals instead of parametric instances of development \cite{nussbaum2003}.} and describe the case of the Wirikuta online movement with the understanding of Appadurai's influential sphere of media\footnote{Appadurai's global analysis of disjunctive flows of people, money, tools, images and ideas describes the influential `scapes' these agents produce and are themselves produced by, and transform communities over time.}; (Sec.3) introduce the normalized web distance (NWD) method  and extend its use to describe temporal  evolution of semantic similarity \cite{cilibrasietal2007}; (Sec.4) describe the Internet experiment, assess the hit-counting using the Google Search engine and the accuracy of the temporal NWD evolution method and (Sec.5) conclude our findings on empowerment measurement for the Wirikuta case.

\section{Social movements on the Internet, the Wirikuta-case.}
During the past decade, the appearance of social media platforms and the increase in their usage shifted not only technological (consumer) trends but also communication and awareness in society at large. Close to literally, `the world'  witnessed  e.g. the uprisings in Tunisia and Egypt in 2010-2011 and with them the rise of empowered sectors of society through the use of social media. It appeared freedom of expression on the Internet and personal spreading of messages through `mass self-communication' media became a new asset to society. \\
According to Castells (2012), for a social movement to form, individuals must connect emotionally to others through a communication process that requires cognitive consonance and effective communication channels \cite{castells2012}. 
The Internet allows emotions, messages on events or informative documents to be spread,   inducing and amplifying cognitive resonance among a global public. This emotion sharing allows the protesters to overcome fear and challenge the powers that they oppose \cite{castells2012}. 
Social movements such as the ones of the Indignados in Spain during 2011-2012, and the global Occupy Wall Street movement in 2011-2012, or the earlier Zapatista `social netwar'  in 1994-1998 \cite{ronfeldtetal1999},   are triggered by sharing the emotions that rise after a meaningful social event in a field of general concern. \\
This active social information networking has far-stretching consequences. E.g. the uncovered global surveillance programs --revealed by Edward Snowden  \cite{Greenwald2013}-- targeting  the private flow of information, highlight the potential  benefits and dangers of chain reactions by individual igniting sparks in the social network cannot be monitored early enough. But also semi-covert targeted engagement in the social information network by external agencies aim to influence these potential  emergent changes; e.g. by providing a messaging service like ZunZuneo in Cuba \cite{zunzuneo2014}.   \\
We focus here however on the potential of indigenous communities --the Huichol community in this case-- to use ICT upon an informed base in alignment with Nussbaum's human development approach (2011) and Appadurai's agential scape of media \cite{nussbaum2011}.\\
About 40000 Huichol, or Wixaritari\footnote{\emph{Wixaritari} is the name of  the people in their own denomination.}, inhabit  scatteredly or in hamlets an area of about  5000 km$^2$ in Western Sierra Madre in the central part of Mexico  \cite{liffman2000}.
Of this region the mayor part is under indigenous control, the remaining part stays culturally closely related.  At the core of the Wixarika\footnote{The name for their language and also the adjective form of the name for their people \emph{Wix\'aritari}.} culture is their sacred land of pilgrimage, the Wirikuta\footnote{A cultural and natural reserve sized about 140000 hectares.}, which extends south of the town of Real de Catorce in the state of San Luis Potos\'i, about four hundred kilometer to the East from their homeland. A detailed anthropological insight in the cultural relation of the Huichol and their pilgrimage to the Cerro Quemado in Wirikuta territory is exposed by Liffman (2000).  Following previous protection measures - like  the Huaxa Manaka Pact \cite{huauxa2008} - by the local authority, the  ``Huichol Route through the sacred sites to Huiricuta"   has been on the tentative list for UNESCO World Heritage since 2004  while the ``Pilgrimage to Wirikuta'' has been proposed as UNESCO Intangible Cultural Heritage by the Mexican government in 2013 \cite{unesco2004,unesco2013,reyna2014}.\footnote{Earlier de facto protection of the Wirikuta  region by UNESCO is not documented. The ``Camino Real de Tierra Adentro'', the ``Historic Centre of Zacatecas'' and the ``Historic Town of Guanajuato and Adjacent Mines" are nearby sites that have received official UNESCO protection - situation in 2013.}\\
In 2010, the Mexican federal government granted open-sky mining concessions to boost economic development in the state of San Luis Potos\'i, a region known for small scale historic mining sites. This area is classified as highly marginal by the government, with a  Human Development Index that fluctuates amongst the lowest in the country  \cite{portalcatorce2014,undp2014}.  500 direct jobs and 1,500 indirect jobs were envisaged along with a planned investment of over 17 millions MXP or 1 million EUR approximately \cite{vilchez2014}.\\
The exploitation however would also take place within Wirikuta, one of the five sacred places for the Huichol community.  By granting exploitation rights to the mining companies, the federal government would renounce the constitutional rights of land, property and self-government by the indigenous group.\footnote{Mexico  endorsed the rights of Tribal and Indigenous Peoples Convention 169 in 1992.} 
In order to prevent the destruction of their cultural sites and natural deterioration from the mining concessions, the Huichol and concerned  supporters organized for protection, they used social media to achieve their pursuit of international acknowledgment mediating their voices, images, sounds and life experiences on the Internet.\footnote{Some of these documents became very popular, like the video ``Wirikuta se defiende! Aho Colectivo"  on YouTube, https://www.youtube.com/watch?v=YQcyxH9q55c. } 
During the second half of 2010 a resistance movement was started to halt  the mining projects: the ``Frente en Defensa de Wirikuta -- Tamatsima Wahaa" \cite{ajagi}. 
There were demonstrations against the governmentÕs actions, but the movement really came to prominence during the second half of 2011 with the support of NGOs.
Information was recorded in Wixaritari and  Spanish, translations in English, French, German and Italian followed.
The spread of information over social media and the response of national and international organisations\footnote{E.g the PEN-international petition letter to president Felipe Calderon, following the \emph{Grupo de los Cien Internacional}'s letter signed by 100 politicians, scholars, writers and artists \cite{grupo2011}.}  and prominent activists \cite{barnett2012a}, influenced for the federal court suspension of February 2012 of the La Luz mining project in Wirikuta until final arbitration.\footnote{Federal court decision of 26/12/2012 concerning  38 concessions in Wirikuta.   http://www.frenteendefensadewirikuta.org/?p=2528\&lang=en} Remaining concessions on the territory were however not affected.  This situations spurred the online activism. A  Wirikuta Fest attracting crowds as large as 60000 to the Foro Sol stadium in Mexico City were organized in 2012  and 2014 \cite{barnett2012b}, short and long films\footnote{We mention e.g. the documentary ``Huicholes: The Last Peyote Guardians" \cite{vilchez2014}} along with 
fashion items\footnote{Including traditional clothing, yarn thread paintings and beaded objects with mythical animal and peyote design} were released to create awareness of the issue and to raise the funds needed to support the movement.\\
In September 2013, with the influence of the movement,  the judicial power of San Luis Potos\'i at the level of the federal court  suspended all remaining mining concessions --Universo and  Maroma with 40 concessions-- in Wirikuta.\footnote{Good News - Mexico: Mining Concessions Suspended in Wirikuta. http://www.culturalsurvival.org/news/good-news-mexico-mining-concessions-suspended-wirikuta}  With this suspension, for now the sacred land of the Huicholes is protected and their traditions sheltered. \\
This short exposition illustrates the various stages of the Wirikuta-case have been documented on the Internet during, or following closely their deployment. The data content of the Internet should thus reflect the evolution of the events, in as much the Internet retains its history. Often new content is written over older content and as such this past information would be lost. Often however the older information is multiplicated to other locations before it is replaced and thus not-updated time-tagged information remains available on the Internet. In the next section we will develop this `tree-ring-growth structure' of the internet and investigate how this model enables a temporal meta-data analysis of the Wirikuta-case.

\section{Temporal evolution of the Normalised Web Distance.} 
Web search engines have become valuable tools to research subjects and evolutions in the electronic, global public sphere.  A simplified approach to evaluate the extent of a case -- not necessarily its presence and priority in the classical online news outlets --  would be the absolute number of  related documents on the Internet. Such could be revealed by counting pages with relevant terms via web search engine. An attestation of this rising impact of the Huichol people's action would be provided by terms such as \emph{Huichol}, \emph{Wixarika} and \emph{Wirikuta}, increasing over time as  counted via e.g. Google Search. 
The temporal evolution from 1994 till 2013 is obtained by restricting for each year Y the time window of the query to 01/01/Y - 12/ 31/Y, and recording the number of page hits for each query.\footnote{Time-tagging of online documents can be ambiguous if upload time-tag and creation time-tag differ greatly.} Each query is thus time restricted to the domain of an annual increment of the Internet. This corresponds to probing consecutive `tree-rings' of the growing Internet. The validity of this approach requires a strong time-tagging of a document which should be correctly identified by the search tool (e.g. \cite{webster2007} for a study using Google Scholar Search). Our choice of probing the internet annualy instead of more frequently will only allow us to track relatively persistent mid-term evolutions. We considered that the modest extent of the Wirikuta-case (as compared to e.g. the Egyptian Uprise), restricted local ICT, would not systematically lead to an extensive fast expression on the Internet. Therefore the time window of one year may average out possible short term variations of NWD.\\
Moreover the size of the Internet, the assynchronous global distribution of the web-indexing data and the commercially protected algorithm of the search tool amounts to  an estimated number  (and time- and user-variable)  instead of  true number of pages \cite{satohetal2012}. In order to estimate these errors, queries were repeated on three different computers at a different moment\footnote{On one computer for single queries measurements were done at three different moments, setting the total at N$_{exp}$=5 for these measurements.}, and the standard error was derived from the standard deviation of the hit counts. Finally the quantity or absolute size does not necessarily reflect unique elements of information, since often content is integrally or partially copied and redistributed to other URL's.
Such a study on temporal evolution of hit counts for terms on a restricted part of the Internet  using Google Scholar has been conveyed previously to map a social scholarly change \cite{webster2007}.

\begin{figure}[!h]  
   \centering
   \includegraphics[width=5in]{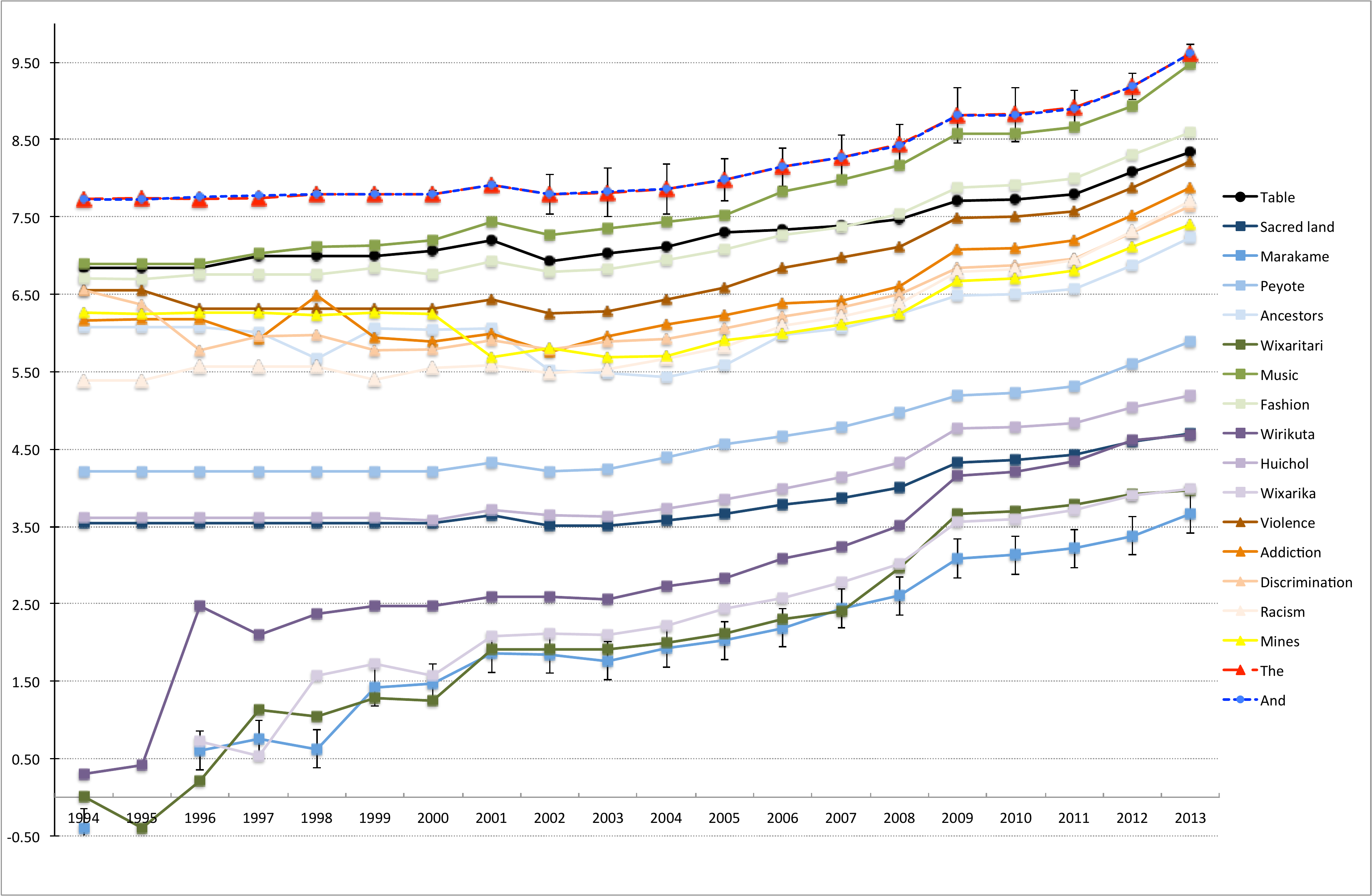} 
   \caption{Annual logarithmic incremental growth of page counts $Log_{10}(n_{year} (u))$ for term $u$. Standard error bars included for highest and lowest measurements indicatively  (N$_{exp}$=5). Systematic shallow `cusp-like' evolutions at measuring points 2001 and 2009 occur. The \emph{the}-curve coincides with the \emph{and}-curve at given resolution. Apparent  linear growth of terms with higher counts prior to 2001 is a Google Search estimation artefact. }
   \label{fig:logincgrowth}
\end{figure}

From the temporally confined Google hit count a sub-exponential (linear) growth till 2004 is followed by an exponential growth for most queries, except the very rare  terms --with counts between 0 and 500-- as \emph{Marakame}\footnote{The Wikarika name for a shaman or traditional spiritual mediator.}, \emph{Wix\'aritari}, \emph{Wixarika} and  \emph{Wirikuta}, which augmented exponentially over the interval except for a linear period of 2002-2003  (see Fig. \ref{fig:logincgrowth}). Clearly no conclusive information can be obtained from  absolute hit count  in temporal evolution: the Google hit estimation algorithm may well increase errors to useless level \cite{kilgarriff2007}. \\
Instead of relying on absolute number estimates we studied the possibility of a semantic method, which is in principle `less' sensitive to absolute size.  The normalized web distance between two terms --as defined in information theory-- gives a measure for their semantic similarity \cite{cilibrasietal2007}. It provides a measure between 0 and 1 which weighs co-occurence, single occurences and Internet size according the amount of Kolmogorovian information which is neccessary to transform  the first term into the other \cite{cilibrasietal2007}.  It has been shown to effectively express semantic similarity.\footnote{E.g. the terms \emph{the} and \emph{and} have very high co-occurence on the Internet and thus have strong semantic similarity and a NWD(the,and) which is approximately zero (see Fig. \ref{fig:logincgrowth}).}
For two terms $u$ and $v$,  with respective hit-counts $n_y(u)$ and $n_y(v)$ and with $n_y ( u \rm{AND} v)$ counts for the search \emph{u} AND \emph{v}, each in the temporal confinement of the year $y$, the normalized web distance $NWD_y(u,v)$ is given by:
\begin{eqnarray}
NWD_y (u,v) &=& \frac{\ln M_y - \ln \mu_y}{\ln N_y - \ln m_y} \label{eq:NWD}
\end{eqnarray}
where $M_y = \rm{max} \left\{ n_y(u), n_y(v) \right\}$, $m_y = \rm{min} \left\{ n_y(u), n_y(v) \right\}$ and $N_y$ is the increment of the internet in the year $y$. The latter number is  --following Vit\'anyi\footnote{Essentially the number $N$ has to be a factor of a few powers larger than $ n_y(u), n_y(v)$ and $n_y ( u \rm{AND} v)$.  ``\dots. N which is the sum of the numbers of occurrences of search terms in each page, summed over all pages indexed. \dots ''.  But, it is stated ``\dots. This parameter N can be adjusted as appropriate, and one can often use the number of indexed pages for N. \dots'' \cite{cilibrasietal2007}}-- a number that is chosen such as to keep the maximum  NWD below 1.\footnote{The NWD is not strictly a distance since it does not necessarily satisfy the triangle inequality. } In our method we must relate $N_y$ to the annual increment, since we relate $u$, $v$ and their co-occurence in that segment of time. We have chosen to use the annual increment of the page count of the term \emph{the}, multiplied by a constant factor set at 100.\footnote{Only two measurements with very low joint counts  $n_{1994}(Huichol\ {\rm AND}\ music)$ and $n_{1994}(Huichol\ {\rm AND}\ fashion)$ remain with NWD  larger than 1 in the first year 1994 of our measurement -- falling in the domain 1994-2000 of discarded data.  } 

In the tree-ring growth-model of the internet, we assume uploaded content in a time frame $y$ to reflect the semantic relation between $u$ and $v$ without important contamination by documents with other time-tags. The NWD should thus reflect some of the internet activism along the timeline of true world events deploying and percolating into the Internet and to some extent, reciprocally, actions undertaken due to mediatised global scale social pressure building.
\begin{figure}[h] 
   \centering
\includegraphics[width=8 cm]{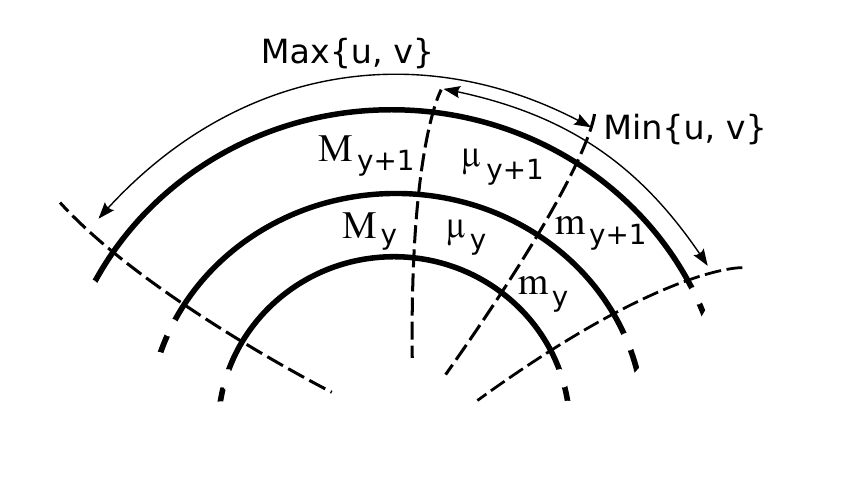} 
   \caption{Scheme of tree-ring growth-model with max, min and joint counts indicated for year $y$ and subsequent year $y+1$. Relative changes $M_y$, $m_y$, $\mu_y$ and $N_y$ in comparison to previous year give rise to change of the $NWD(u,v)$. }
   \label{fig:TreeRing}
\end{figure}
We could expect classifier terms to change their NWD with respect to some central key terms identifying the social activism case.
The temporal variation of the NWD depends on the changing presence of the terms $u$, $v$ and $u AND v$ in the annual increment $N$ of the Internet.   
\begin{eqnarray}
\delta NWD (u,v) 
& =&   \frac{1}{ \ln N - \ln m} \left(  \frac{\delta M}{M} - \frac{\delta \mu}{\mu}  +  NWD (u, v)   \left(  \frac{\delta m}{m} - \frac{\delta N}{N} \right)  \right) \label{eq:dNWD}
\end{eqnarray}
with $\delta M = M_{y+1}-M_y$, idem for the other quantities.  In the event of an observed change of NWD, it is possible to inspect what precisely its origins are.
In our case of study, three key terms were identified for the `Wirikuta-case' and thirteen classifiers were chosen and subdivided in three (fuzzy) categories; positive, negative and neutral.
The list of positive classifiers  includes terms  that would normally be related positively to culture,  hence determined as indicative --or with importance-- for cultural heritage and identity by Huichol people. 
The list of the negative classifiers contains terms that could be perceived to have a negative relation to the Huichol culture.  Evidently this semantic classification is not static i) the syntax of a sentence in which a term occurs will to large extent determine its connotative meaning, ii) metonymic use of terms like \emph{sacred land} or \emph{music} in language,  iii) polysemy of terms, e.g. \emph{fashion} will mean both `a manner of doing something' or `a trend of style', and iv) the language mixing Wixarica terms are lent to Spanish and lent to English which leads to measuring NWD between terms of a different languages.\\
One should take into account the possible issue of  measuring the Internet in a `filter bubble' \cite{pariser2011}. Procedures of commercial optimization, user predilection adapting and institutional censorship could modify the ranking but also quantity of query results. The NWD expression Eq. (\ref{eq:NWD}) is not sensitive to ranking but evidently to selective modification of $M_y$, $m_y$, $\mu_y$ and $N_y$.  Only homogenous scaling of the parameters $M_y$, $m_y$, $\mu_y$ and $N_y$ leaves the NWD invariant.\\
 In order to control our interpretation of  approaching or receding NWD over time, we considered a term that would maintain a neutral relation with respect to the Huichol culture, and chose the term \emph{table} as a common piece of household furniture.\footnote{It is almost impossible to chose a term  which does not `suffer' from too much polysemy; \emph{table} can stand for `piece of furniture', `a format to display numbers' or  `a flat surface', all of which are mostly considered as neutral.  E.g. \emph{mines} can be used for `explosive devices' or `installations for mineral extraction' which tend to be perceived as negative or neutral. }  The invariability of this neutral term NWD with respect to a key term would indicate the quality of the proposed method and possible artifacts of the search tool. \\
 Finally a of temporal evolution of  the NWD of two common English terms \emph{the} and \emph{and} was made as well in order to control the control distance based on \emph{table}.

\section{Methodology of the Internet  measurement experiments}
For the three key terms  central to the `Wirikuta case' we have chosen  \emph{Wirikuta}, \emph{Huichol} and\emph{Wixarica}.  The seven  selected positive classifiers are; \emph{sacred land}, \emph{Marakame}, \emph{peyote}, \emph{ancestors}, \emph{Wixaritari}, \emph{music}, \emph{fashion}. The five selected negative classifiers are; \emph{violence}, \emph{addiction}, \emph{discrimination}, \emph{racism} and  \emph{mines}. The single neutral classifier was chosen to be \emph{table}. 
The measurements consist of manually executing  queries on a freely available search engine --here Google Search-- which provides an estimate of the hit counts and which allows a `Custom date range' functionality to a query. The date range was chosen to coincide with the calendar year  01/01/Y - 31/12/Y.  
 Starting with  $Y=1994$ up till $Y=2013$ the hit counts of all queries were recorded. \\  
In order to calculate the ${\rm NWD} (u,v)$ also joint queries of key terms $u$ with classifier terms $v$ were made. The query terms always were embedded in quotation marks --``term''--  in order to avoid stemming or associative query results which tend to be provided by search engines. For joint queries the Boolean operator AND was added in the query: ``u" AND ``v''.   The key terms were also checked for NWD temporal variation among them.\\
Considering the variability of Google's estimation algorithm for hit counts, we worked on three different computers, from different IP addresses. The hit counts for the queries were retrieved in the period 17/02/2014-24/02/2014, in the Brussels  region in Belgium.\footnote{ Two missing joint query series were added in the period 18/04/2014-24/04/2014, the numbers where in line with previous measurement of February and were added to the data set.}
The standard error on the average measurements was obtained using the standard deviation with  N$_{exp}$=5 for single queries and N$_{exp}$=3 for joint queries.\footnote{The error on each hit number seems to be set by Google to three significant numbers. This sometimes leads to quixotic outcomes in the lower range spectrum. It is in general a vast underestimation of the accuracy. }

\section{Interpretation of results of web-measurements and conclusion.}
The growth of the internet is clearly apparent from the `date range' queries; the tree-ring growth shows an exponential increase after 2001 (near linear tendency  in logarithmic scaled Fig. \ref{fig:logincgrowth}).\footnote{The data set of single and joint counts is available by email to jbroekae@vub.ac.be. } 
From absolute counts  we notice exponential growth on \emph{Wirikuta}, \emph{Wixaritari}, \emph{Wixarika} and \emph{Marakame} over the full time period of our measurement 1994-2013: all of them rare Wixarika language terms  which  start at zero or almost zero counts.  The  term  \emph{Huichol}   --closely related to Wixarika-- starts at higher counts in 1994  and does  not expose this exponential growth but until 2001, identically as all terms with higher initial counts do. Given the possibility of artefacts in the count estimate algorithm of Google Search it is not possible to confer any interpretation to this difference in growth prior to 2001.\footnote{We noticed a nearly systematic estimation effect of hit counts by Google Search in date range modus for queries from 1994 till 2000 for terms with counts above approximately 100. For example the term \emph{peyote} receives for seven consecutive annual measurements exactly the same number 16200 (measured on 16/02/2014). } These artefacts disappear starting 2001 and led us to discard any interpretation of data with counts higher than approximately 100 in the period 1994-2000. \\
Of all eighteen terms measured -- key, classifier, control -- the two fastest growing terms were \emph{Wixaritari} with slope $0.21 \pm 0.02$ ($R^2=0.91$) and  \emph{Wirikuta} with slope $0.20 \pm 0.02$ ($R^2=0.94$).  The slowest growing term was \emph{table}, $0.10 \pm 0.01$ ($R^2=0.90$).\\

%

\begin{figure}[htb]  
   \centering
   \includegraphics[width=5in]{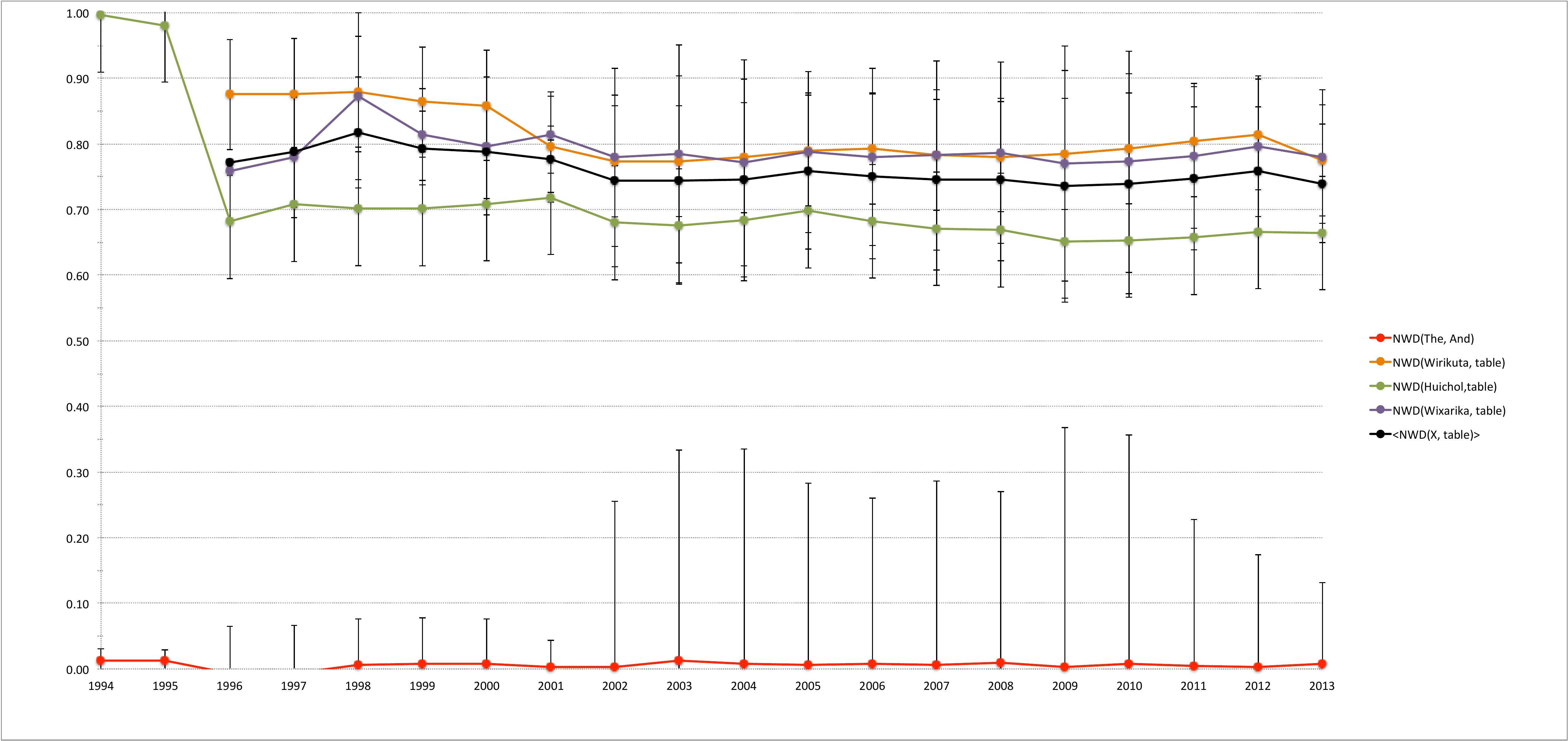} 
   \caption{Annual change of control NWD's; for \emph{the} and \emph{and} near zero, and for \emph{table} with respect to \emph{Wirikuta} ($\approx 0.79\pm0.13$), \emph{Huichol} ($\approx 0.67\pm0.12$) and \emph{Wixarika} ($\approx 0.78 \pm 0.13$). A small systematic co-variation of the NWD is apparent starting measuring point 2001. $<$NWD(u, table)$>$ is the average of the three distances relative to the key terms $u$ and is used for assessing the control evolution ($\approx 0.75\pm0.13$).}
   \label{fig:controlNWD}
\end{figure}

\begin{figure}[htb]  
   \centering
   \includegraphics[width=5in]{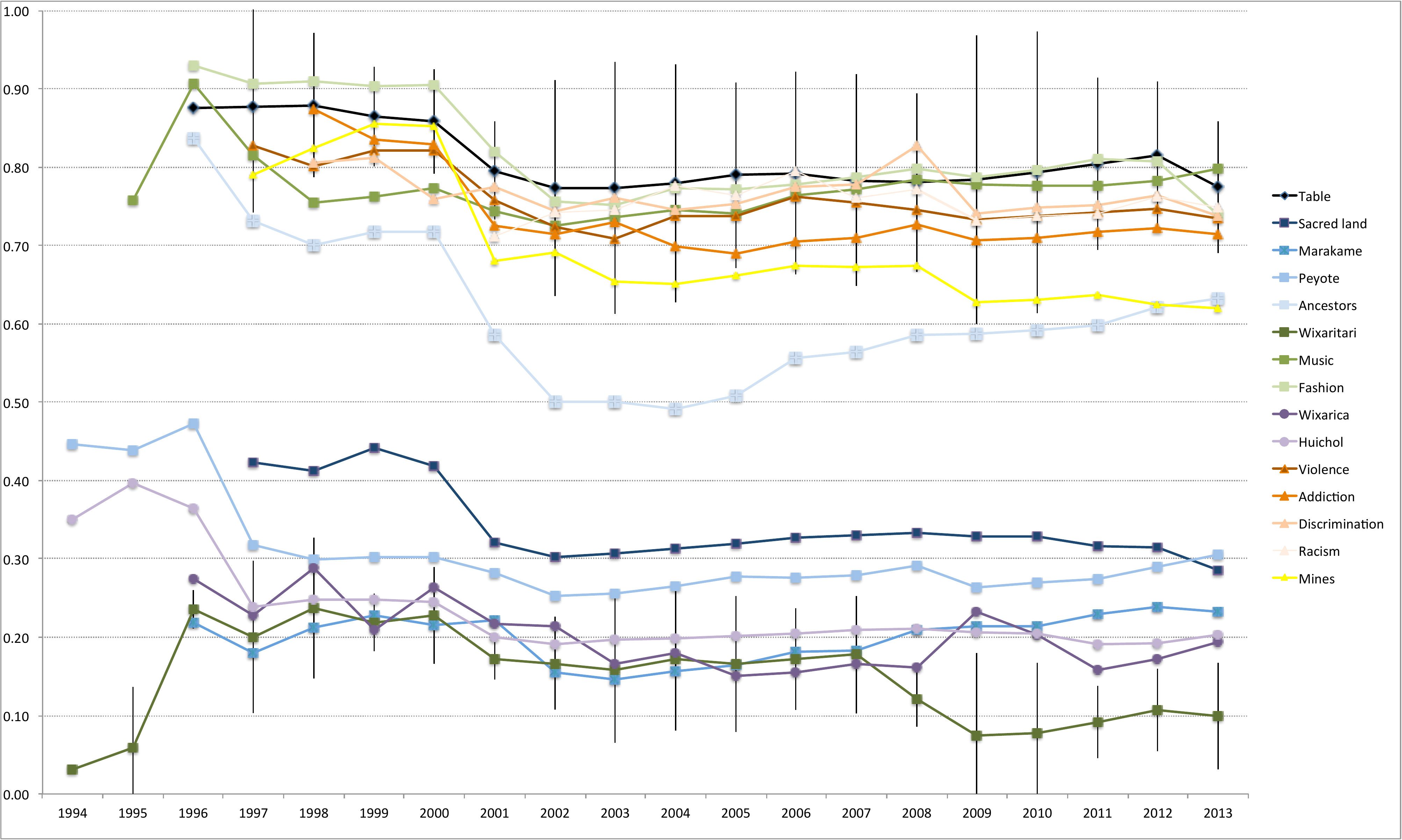} 
   \caption{ Temporal evolution of  NWD(Wirikuta, u).  Valid NWD  variation is considered starting at measuring point 2001. }
   \label{fig:wirikutaNWD}
\end{figure}

\begin{figure}[htb]  
   \centering
   \includegraphics[width=5in]{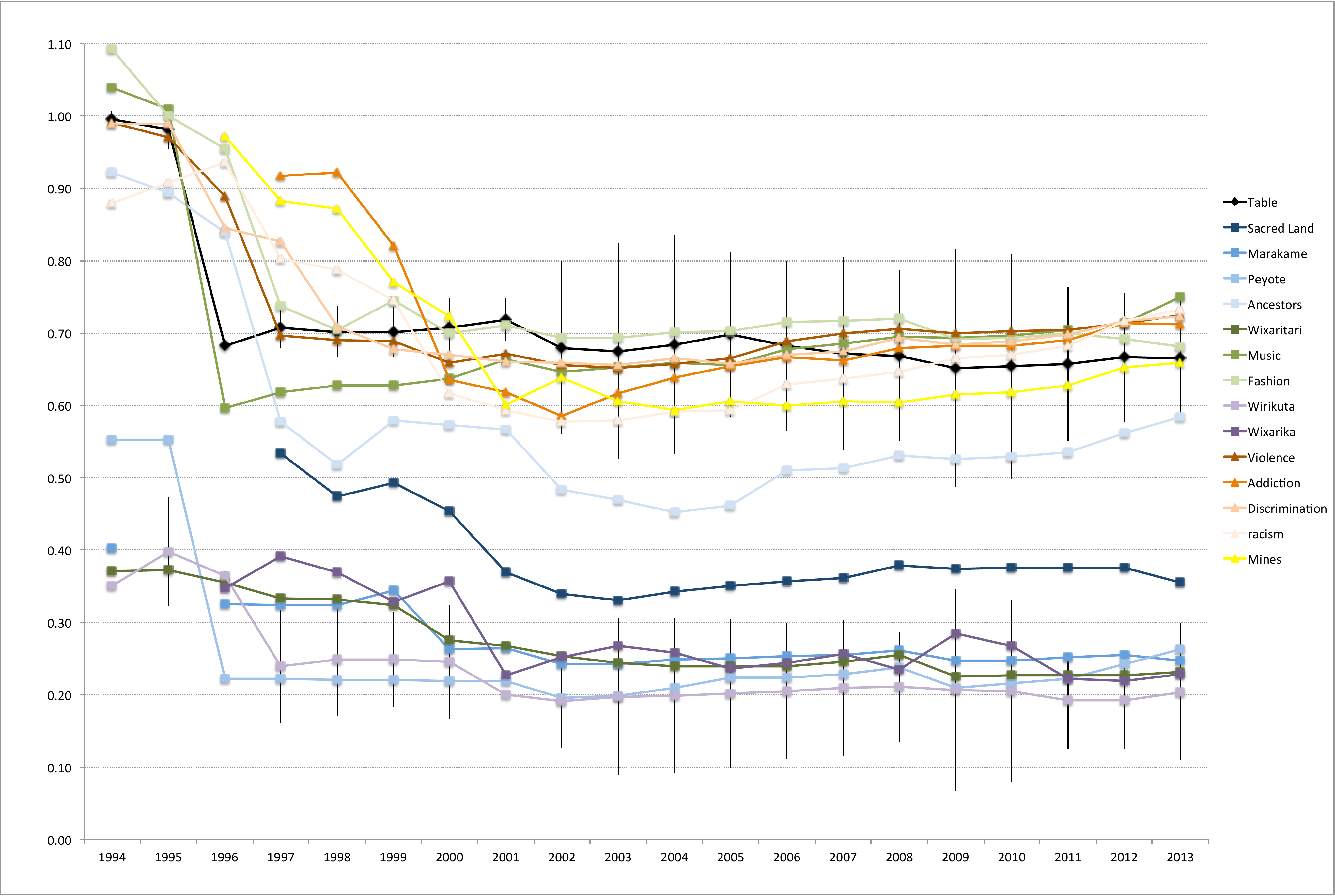} 
   \caption{ Temporal evolution of  NWD(Huichol, u). Valid NWD  variation is considered starting at measuring point 2001. }
   \label{fig:huicholNWD}
\end{figure}

\begin{figure}[htb] 
   \centering
   \includegraphics[width=5in]{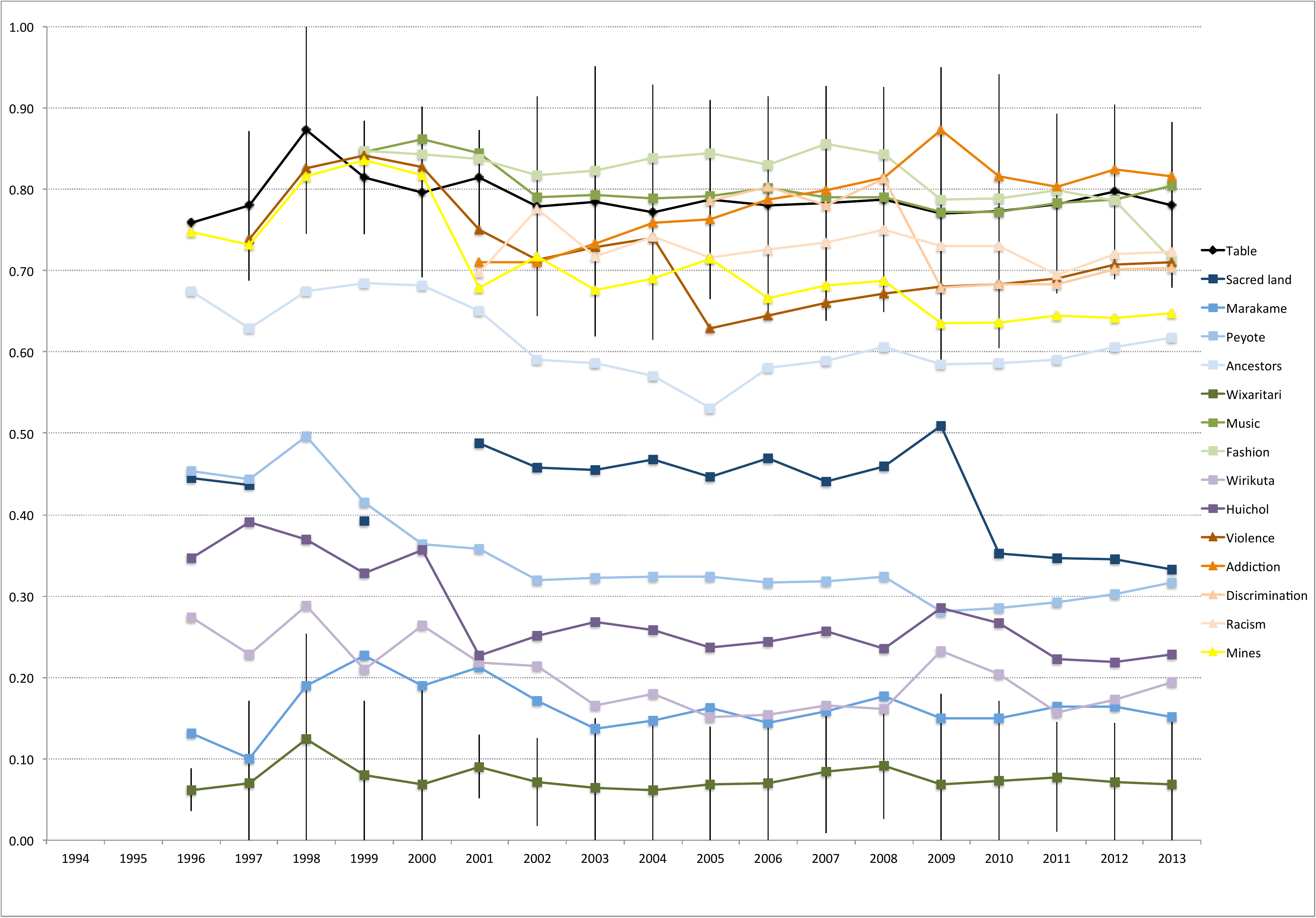} 
   \caption{ Temporal evolution of  NWD(Wixarika, u). Valid NWD  variation is considered starting at measuring point 2001. }
   \label{fig:wixarikaNWD}
\end{figure}

The NWD requires  counts for joint queries using the Boolean connector AND. However on examination Google  Search returns hit counts that  e.g. do not satisfy the inclusion-exclusion principle.\footnote{The principle would require: $n_{year} (u \  {\rm OR}\ v) = n_{year} (u) + n_{year} (v) - n_{year} (u \ {\rm AND}\ v) $} Therefore one should critically assess as well the  Boolean conjunction in queries.\\
First we observe the NWD adequately expresses semantic similarity in the sense that small distance relate \emph{Wirikuta} with --in increasing order in 2013-- \emph{Wixaritari} ($0.10\pm0.07$), 
\emph{Wixarica} ($0.19\pm0.04$), 
\emph{Huichol} ($0.20 \pm 0.09$), 
\emph{Marakame} ($0.23\pm 0.07$), 
\emph{sacred land} ($0.29\pm0.08$) and 
\emph{peyote} ($0.31\pm 0.10$). 
Larger distances are noticed for e.g.\emph{mines} ($0.62\pm0.07$), \emph{ancestors} ($0.63\pm0.09$),  and all other negative classifiers around  0.70-0.75 as well as \emph{music} and \emph{fashion} 0.75-0.80 and the control classifier \emph{table} ($0.77\pm0.08$).\\
For the term \emph{Huichol} we notice the nearness of the same terms 
\emph{Wixaritari} ($0.23\pm0.09$), 
\emph{Wixarica} ($0.23\pm0.04$), 
\emph{Wirikuta} ($0.20 \pm 0.09$), 
\emph{Marakame} ($0.25\pm 0.07$ ), 
\emph{sacred land} ($0.36\pm0.09$)+ and 
\emph{peyote} ($0.26\pm 0.11$). 
More remote terms contain again the negative classifiers (range 0.65-0.75) --e.g \emph{mines} ($0.66\pm0.10$)--  \emph{music} ($0.75\pm0.10$), \emph{fashion} ($0.68\pm0.09$) and the control term ($0.66\pm0.09$).
Finally for the term \emph{Wixarika} we notice again the nearness of the same terms
\emph{Wixaritari} ($0.07\pm0.08$), 
\emph{Wirikuta} ($0.19 \pm 0.04$),
\emph{Huichol} ($0.23 \pm 0.04$),
 \emph{Marakame} ($0.15\pm 0.07$ ), 
 \emph{sacred land} ($0.33\pm0.09$) and 
 \emph{peyote} ($0.32\pm 0.10$).  More remote terms contain again the same negative classifiers (range 0.70-0.75) --e.g \emph{mines} ($0.65\pm0.09$) and, \emph{music} ($0.80\pm0.09$), \emph{fashion} ($0.71\pm0.11$) and the control term ($0.78\pm0.10$).\\
Closest NWD are thus found among terms of the same language   \emph{Wirikuta},  \emph{Wixarika} and \emph{Wixaritari}. The term \emph{sacred land} is marginally more related to the land \emph{Wirikuta} and, \emph{Marakame} is closer to the people named in their own language \emph{Wixarika} than in Spanish \emph{Huichol}. Notice here that \emph{mines} has about the same distance to all three key terms (measurement 2013).\\
The  evolution of the control $\left< \rm{NWD}_y (\emph{table}) \right>$  where the average has been taken over the key terms  and its control NWD$_y$(\emph{the},\emph{and}), show a constant value starting 2002 (Fig. \ref{fig:controlNWD}). We recall that due to artefacts in returned count estimates by Google Search we discard interpretation of data prior to 2001.
The linear approximation of the control NWD evolution of \emph{table} with respect to  \emph{Wirikuta}  has a slope $0,001\pm 0,001$ ($R^2=0,148$), for \emph{Huichol}  slope $-0,0036 \pm 0,001$ ($R^2= 0,579$) and for \emph{Wixarika} slope $-0,001 \pm 0,001$ ($R^2=0,072$); on average we find for the NWD of \emph{table} to the key terms the slope is $-0,001\pm 0,001$ ($R^2=0,153$).  The control assessment by use of the invariant semantic relation of \emph{the} and \emph{and} by NWD(\emph{the}, \emph{and}) with slope $-0,0001\pm0,0002\ (R^2=0,0057)$ shows therefore strong alignment with $\left< \rm{NWD}_y (\emph{table}) \right>$ with however a slightly larger variability.\footnote{For invariant time evolution --a horizontal line-- the coefficient of determination $R^2$ by definition gives no qualitative indication.}
 We conclude that our neutral  term \emph{table} thus indeed exposes no significant change of semantic similarity with the key terms, as we had hypothesized. This entitles to certain extent to meaningfully interpret NWD temporal evolutions of terms after 2001 using this method. However given the standard error due to the variability of the counts we will adopt the rule that at least a slope of $\pm 0.005$ is necessary to hint at a change of semantic similarity, while the relative error should be not more than $\frac{1}{3}$.\footnote{The interval of 12 years then leads to an approximate change of 0.06 in NWD.} Evolutions with a lesser slope wil be considered constant over time (at this time resolution). \\
First we notice in general a lesser variability among NWD of \emph{Wirikuta}  (Fig. \ref{fig:wirikutaNWD}) and \emph{Huichol} (Fig. \ref{fig:huicholNWD}) as compared to \emph{Wixarika} (Fig. \ref{fig:wixarikaNWD}).  The lesser count rates for the latter term leads to  larger fluctuations of the NWD.\\
The observed rather systematic plunge of the NWD evolution from the start of our measurements in 1994 to 2001-2002  in the three graphs (Figs. \ref{fig:wirikutaNWD},  \ref{fig:huicholNWD}, \ref{fig:wixarikaNWD}) can be retraced to anomalous slow growth of the $\delta M$ (where $M=Max\{u,v\})$ in Eq. (\ref{eq:dNWD}). Again this effect can be reduced essentially to artefactual returned count estimates by Google Search prior to 2001.\\
We will therefore only assess changes of NWD starting 2001, again the high variability of the counts allows for a linear regression at most. \\
For \emph{Wirikuta} (Fig. \ref{fig:wirikutaNWD}) we observe the receding of the terms  \emph{ancestors}  ($0.010\pm0.002, R^2=0.616$), \emph{Marakame} ($0.006\pm0.002, R^2=0.532$)  and \emph{music} ($0.005\pm0.001, R^2=0.841$)  and the approaching of \emph{Wixaritari} ($- 0.008\pm0.002, R^2=0.650$) and \emph{mines} ($- 0.005\pm0.001, R^2=0.643$).\\
For \emph{Huichol} we observe the following receding terms \emph{racism} ($0.0130\pm0.001 , R^2=0.933$), \emph{music} ($0.007\pm0.001 , R^2=0.851$), \emph{violence} ($0.006\pm0.001, R^2=0.835$) and \emph{discrimination} ($0.005\pm0.001 , R^2=0.873$), no terms are approaching.\\
For \emph{Wixarika} we observe one receding term \emph{addiction} ($0.010\pm0.002, R^2=0.723$) and the approaching terms \emph{discrimination} ($-0.016\pm0.005, R^2=0.591$), \emph{sacred land} ($-0.012\pm0.003, R^2=0.570$), \emph{fashion} ($-0.007\pm0.002, R^2=0.489$) and \emph{mines} ($-0.005\pm0.001, R^2=0.578$).

From the analysis of the graphs of NWD-evolution and the standard errors on them we find a weak indication of semantic change and overal a complex evolution  for the classifiers with respect to the key terms. Within our set of classifiers the territory \emph{Wirikuta} has slightly gained relation with their proper named people \emph{Wixaritari} and its disputed \emph{mines} and, slightly lost relation with the English term \emph{ancestors} and the name for the shaman \emph{Marakame}. The term \emph{Huichol} has taken slightly more distance from a number of negative classifiers, but also from \emph{music}.\\
Finally \emph{Wixarica} has taken a closer relation to the negative terms \emph{discrimination} and \emph{mines} but also to \emph{fashion} and \emph{sacred land} and had slightly less relation to the term \emph{addiction}.\\
We observe in the NWD-evolution therefore a reflection of the complex engagement of the Huichol with the mining threat on their land. The present analysis does however not allow a better resolution of event development and its representation in the public sphere of the Internet. A more reliable tool for precise hit counting,  precise time-labeling and more frequent measurements over the time-range of the study  should allow to reach more detailed understanding of the dynamics of the Internet and the usage of ICT by communities and their activist expressions. \\
  A future repetition of our 1994-2013 measurements could verify whether the present findings are stable over time and  as such assess the tree-ring growth-model of the Internet.  The main result at this point is the observation that the NWD-evolution on the Internet of key terms assigned to a case of social engagement  does evolve with the events deploying in this case.\\

\section{Annex: NWD-graphs }

\end{document}